\documentclass[twocolumn,prl,superscriptaddress,floatfix,showpacs]{revtex4-1}
\usepackage[utf8]{inputenc}
\usepackage{amsmath}
\usepackage{amssymb}
\usepackage{graphicx}
\usepackage{multirow}
\usepackage{tabularx}

\makeatletter
\usepackage{graphics}
\usepackage{epsfig}
\usepackage{color}

\newcommand{\be}{\begin{equation}} \newcommand{\ee}{\end{equation}}
\newcommand{\bea}{\begin{eqnarray}} \newcommand{\eea}{\end{eqnarray}}

\makeatother

\begin{document}

\newcolumntype{L}[1]{>{\raggedright\arraybackslash}p{#1}}
\newcolumntype{C}[1]{>{\centering\arraybackslash}p{#1}}
\newcolumntype{R}[1]{>{\raggedleft\arraybackslash}p{#1}}

\title{Collective irregular dynamics in balanced networks of leaky integrate-and-fire neurons}

\author{Antonio Politi}
\affiliation{Institute for Complex Systems and Mathematical Biology and Department of Physics (SUPA), 
Old Aberdeen, Aberdeen AB24 3UE, UK}
\author{Ekkehard Ullner} 
\affiliation{Institute for Complex Systems and Mathematical Biology and Department of Physics (SUPA), 
Old Aberdeen, Aberdeen AB24 3UE, UK}
\author{Alessandro Torcini} 
\affiliation{Laboratoire de Physique Th\'eorique et Mod\'elisation, Universit\'e de Cergy-Pontoise, CNRS, UMR 8089, 95302 Cergy-Pontoise cedex, France}
\affiliation{CNR - Consiglio Nazionale delle Ricerche - Istituto dei Sistemi Complessi, via Madonna del Piano 10, I-50019 Sesto Fiorentino, Italy}

\date{\today}

\begin{abstract}
We extensively explore networks of weakly unbalanced, leaky integrate-and-fire 
(LIF) neurons for different coupling strength, connectivity, and by varying the degree
of refractoriness, as well as the delay in the spike transmission.
We find that the neural network does not only exhibit a microscopic (single-neuron)
stochastic-like evolution, but also a collective irregular dynamics (CID).
Our analysis is based on the computation of a suitable order parameter, typically
used to characterize synchronization phenomena and on a detailed scaling analysis
(i.e. simulations of different network sizes).
As a result, we can conclude that CID is a true thermodynamic phase, intrinsically
different from the standard asynchronous regime.

\end{abstract}

\maketitle

\section{Introduction} 
The use of simple models proves often very helpful to identify and characterize the mechanisms underlying general dynamical phenomena. Computational neuroscience is a field where this approach is potentially
very powerful, given the myriad of interactions involved in the functioning of the mammalian brain \cite{gerstner,dayan2003}. However, setting the appropriate level of simplicity is not a priori obvious. A particularly enlightening example is the reproduction of the background neural activity. Most of the numerical and theoretical
studies are based on the so-called rate models, where each neuron is characterized by a single coarse-grained variable representing the strength of the ongoing activity
\cite{deco2008,ostojic2011}. However, it is well known that neurons work by emitting single spikes, so that it is more natural to represent them as (nonlinear) oscillators. This is, indeed the philosophy adopted by many studies based on pulse coupled units, such as leaky integrate-and-fire (LIF) neurons \cite{burkitt2006}.
Accordingly, a general question arises as to whether the two approaches are consistent with one another and, in particular, to what extent spiking neurons reproduce the scenario observed in rate models \cite{ermentrout2010,montbrio2015}.

In this paper we look at the evolution of the so-called balanced networks, where excitatory and inhibitory 
interactions compensate each other \cite{vogels2005}, since the single-neuron dynamics
is rather irregular and reminiscent of the background neural activity.
A detailed theory of this regime has been developed for rate models in the limit of large 
connectivity \cite{bal1,bal2,bal3,bal4}, but some theoretical studies have been also made for spiking LIF neurons 
in highly diluited networks~\cite{brunel2000}. 
Altogether, the scenario which emerges from these studies is that of an asynchronous regime, i.e. a dynamical state characterized by ``microscopic" fluctuations, but a steady and constant firing rate at the macroscopic level (in the thermodynamic limit). 
Nevertheless, evidence of irregular collective dynamics has been recently found in both rate models \cite{hayakawa2017} 
and spiking neurons \cite{noi}, so that a question arises about the conditions for the emergence of partial synchronization.

In this paper we focus on networks of LIF spiking neurons, as we believe that this is a more realistic and meaningful setup.
We extend the preliminary analysis presented in~\cite{noi}, by including a study of the avalanches and of the scaling 
behavior of the fluctuations of the spiking activity. Furthermore, we explore several variants of the model to test
the robustness of the results of our claims.
More precisely, we refer to a model that has been repeatedly investigated in the literature and in particular 
in~\cite{brunel2000,ostojic} (several parameter values are herein selected as in Ref.~\cite{ostojic}).
However, there are important differences that should be stressed and, which allow us drawing convincing conclusions
about the existence of a CID:
(i) instead of sparse networks, we consider massive ones (i.e. the connectivity $K$ is assumed to be proportional 
to the network size $N$); (ii) the coupling strength is chosen to be of the order of $1/\sqrt{K}$. 
Furthermore, (iii) the unbalance is chosen to be of the same order as the statistical fluctuations of the input current, 
so as to avoid a complete dominance of either excitation, or inhibition. 
Equipped with such assumptions, we have simulated different network sizes, finding that the seemingly CID observed 
for finite networks survives in the thermodynamic limit, thereby validating its identification as of a true thermodynamic phase.

In section~\ref{sec:model} we introduce the model, briefly review the numerical scheme adopted for its simulations, and discuss a problem related to its ill-defined structure. In fact, in this model strictly simultaneous events can unavoidably occur, which require an additional protocol to specify the way they should be treated. This phenomenon is related to the occurrence of avalanches which we show to marginally influence the thermodynamic limit.
In section~\ref{sec:micro}, we deal with the dynamical properties of the single-neuron dynamics, mostly focusing on the statistics of the inter-spike intervals (ISI) and on the spectral properties of the spike trains 
emitted by a single neuron
as well as of the post-synaptic input currents received by each neuron.
In section~\ref{sec:col}, we discuss the collective dynamics, introducing a suitable order parameter to quantify the degree of synchronization, and characterizing its stochastic-like behavior by means of the power spectrum of the global neural activity. The perturbative approach developed by Brunel in \cite{brunel2000} is implemented to perform a comparison with the numerical observations. 
A qualitative agreement is found.

Finally a fractal-dimension analysis is performed, which confirms the stochastic-like character of the dynamics: i.e. its high-dimensional nature.
In section ~\ref{sec:robust}, we explore the collective behaviour of the LIF model when some of the parameters are varied, notably, delay, refractoriness, connectivity and the presence of external noise. All of the simulations confirm the robustness of CID. Section~\ref{sec:conclusion} contains a summary of the main results and a brief list of the main open problems.

\section{The model} \label{sec:model} 

In this Section we define the model following Ref.~\cite{brunel2000,ostojic}, equipped
with a suitable scaling of some parameters, in order to preserve the CID observed at finite sizes.
Furthermore, we discuss some intrinsic ambiguities present in the definition of the model
that are related to the unavoidable occurrence of strictly synchronous events 
(the simultaneous arrival of spikes emitted by different presynaptic neurons crossing the 
threshold at the same moment). The treatment of these events is somehow arbitrary and 
requires the definition of a specific protocol. Furthermore, these synchronous events 
propagate in the network as a sequence of events separated by the propagation-time of the 
single spikes. The resulting avalanches are analysed in detail for networks of increasing
size.

We consider networks of spiking neurons composed of $N$ supra-threshold LIF neurons,
split in $bN$ excitatory cells and in $(1-b)N$ inhibitory ones~\cite{brunel2000,ostojic}. 
The membrane potential $V_i$ of the generic $i$-th neuron evolves according to
\begin{equation}
\tau \dot V_i = R (I_{0}+I_i) - V_i  \; ,
\label{eq:LIF}
\end{equation}
where $\tau = 20$ ms is the membrane time constant, $R I_{0} = 24$ mV is a constant external DC ``current",
and $R I_i$ corresponds to the synaptic current due to the recurrent connections within
the network, namely
\begin{equation}
RI_i = \tau \sum_n G_{ij(n)} \delta(t-t_{j(n)}-\tau_d) \; ,
\label{eq:general}
\end{equation}
where $j(n)$ is the label of the node firing the $n$-th spike.
The synaptic connections among the neurons are random without
autapses, but with a fixed in-degree $K$ for each neuron. 
In particular, we consider a massive network, where the in-degree grows proportionally to the size as 
$K=cN$; the proportionality parameter $c$ is termed connectivity -- unless stated otherwise we set
$c=0.1$ throughout the entire paper.
The random connections are embodied in the matrix $G$, whose elements take the values
 $G_{ij}=J_e$ ($-J_i$), if the pre-synaptic neuron 
$j$ is excitatory (inhibitory), otherwise $G_{ij}=0$.
Whenever at time $t_{j(n)}$ the membrane potential $V_j$ of the $j$-th neuron reaches 
the threshold $V_{th} = 20$ mV for the $n$-th time, two events
are triggered: (i) the membrane potential is reset to $V_r = 10$ mV and it is then held fixed 
for a refractory period $\tau_r=0.5$ ms; (ii) a spike is emitted and received $\tau_d = 0.55$ ms 
later by the post-synaptic cells connected to neuron $j$. 

In order to maintain a fixed balance between excitation and inhibition irrespective of the in-degree
(and, thereby, of the system size) we assume that the coupling strength scales as the inverse of 
the square root of the in-degree, as done in most of the literature on the balanced 
state~\cite{bal1, bal2,bal3,bal4}. More precisely, we assume
$J_e = J\sqrt{1000/K}$ and $J_i = (4 + g_1\sqrt{c/K})J_e$~\cite{noi}.
With these choices, and for $g_1=100$, we recover the setup studied in ~\cite{ostojic} for $N=10,000$, 
$K=1000$ and $b=0.8$. The coupling strength $J$ is our main control parameter.

\subsection{Integration scheme}

The model equations~(\ref{eq:LIF},\ref{eq:general}) can be solved by either implementing an
event driven integration scheme, such as described in~\cite{zillmer2006,klinshov2017chaos}, or a more standard
clock-driven strategy~\cite{integration}. The former scheme is, in principle, exact;
provided that the integration time step is small enough, also the latter scheme is sufficiently
accurate.
In fact, the results discussed in this paper have been obtained by implementing either scheme
without any specific preference. 
One exception is the above mentioned presence of ambiguities in the very definition of the
model, which can be singled out only with reference to the exact event-driven scheme.
In general, two types of event break the smooth evolution of the membrane potential:
(i) a neuron reaches the threshold; (ii) a neuron receives one (or more) post-synaptic potentials (PSPs),
elicited by one or more pre-synaptic neurons $\tau_d = 0.55$ ms earlier. In between these events 
all neurons evolve as being uncoupled according Eq.~(\ref{eq:LIF}) with $I_i=0$. 
In order to evolve the system, we first need to identify the next type of event
and thereby update all the membrane potentials $\{V_i\}$ until the occurrence of the event itself.
This is done by solving analytically Eq.~(\ref{eq:LIF}) with $I_i=0$. 
Thereafter, we process either the threshold passing (i) or the spike-receiving (ii) event. 
Each emitted spike is received $\tau_d = 0.55$ ms later by $K=cN$ other neurons. 
If the sending neuron $j$ is excitatory, it may happen that the post-synaptic potentials
triggers several threshold passings events at exactly the same time. 
Since the delay is the same for all synaptic connections, in the next round,
more than one of the simultaneous spikes can reach the same neuron.
It is, therefore, necessary to complement the definition of the model with a rule to
handle perfectly synchronous spikes, since the outcome depends on the way such events are
treated.

We have decided that the most ``neutral" rule consists in first estimating the net effect of all the PSPs on the basis of the current value $V_i$ of the membrane potential.
Afterwards, all neurons which passed threshold are reset to the common value $V_r$ and the emitted
spikes are stored to be later received by the postsynaptic neurons connected to the firing ones. 
We have verified that also in the case of event driven integration
schemes spike avalanches separated by exactly one time delay can occur. In the next section we discuss how to quantify their relevance.

\subsection{Characterization of the avalanches}

\begin{figure}
\begin{centering}
\includegraphics[width=0.44\textwidth,clip=true]{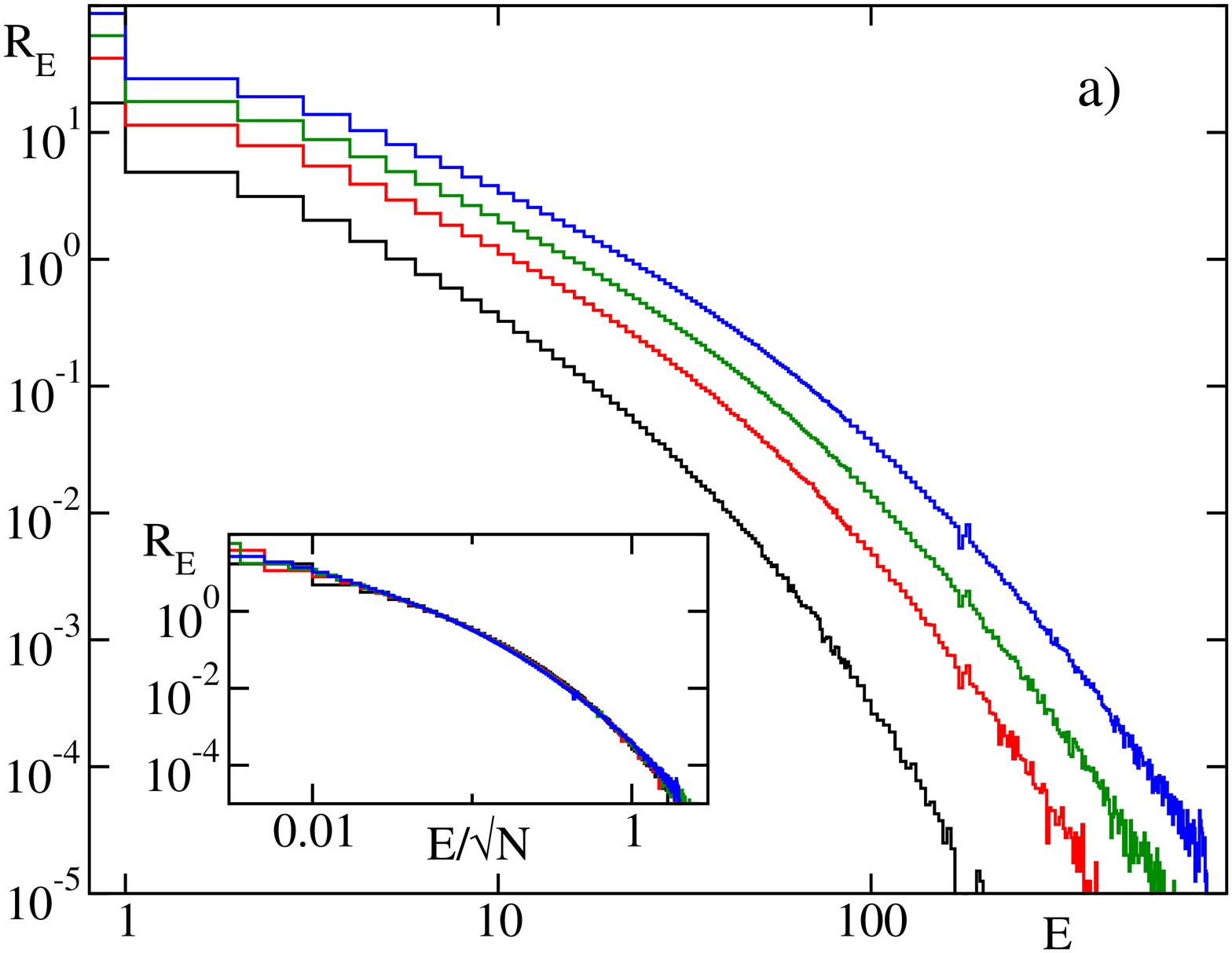}\\
\includegraphics[width=0.44\textwidth,clip=true]{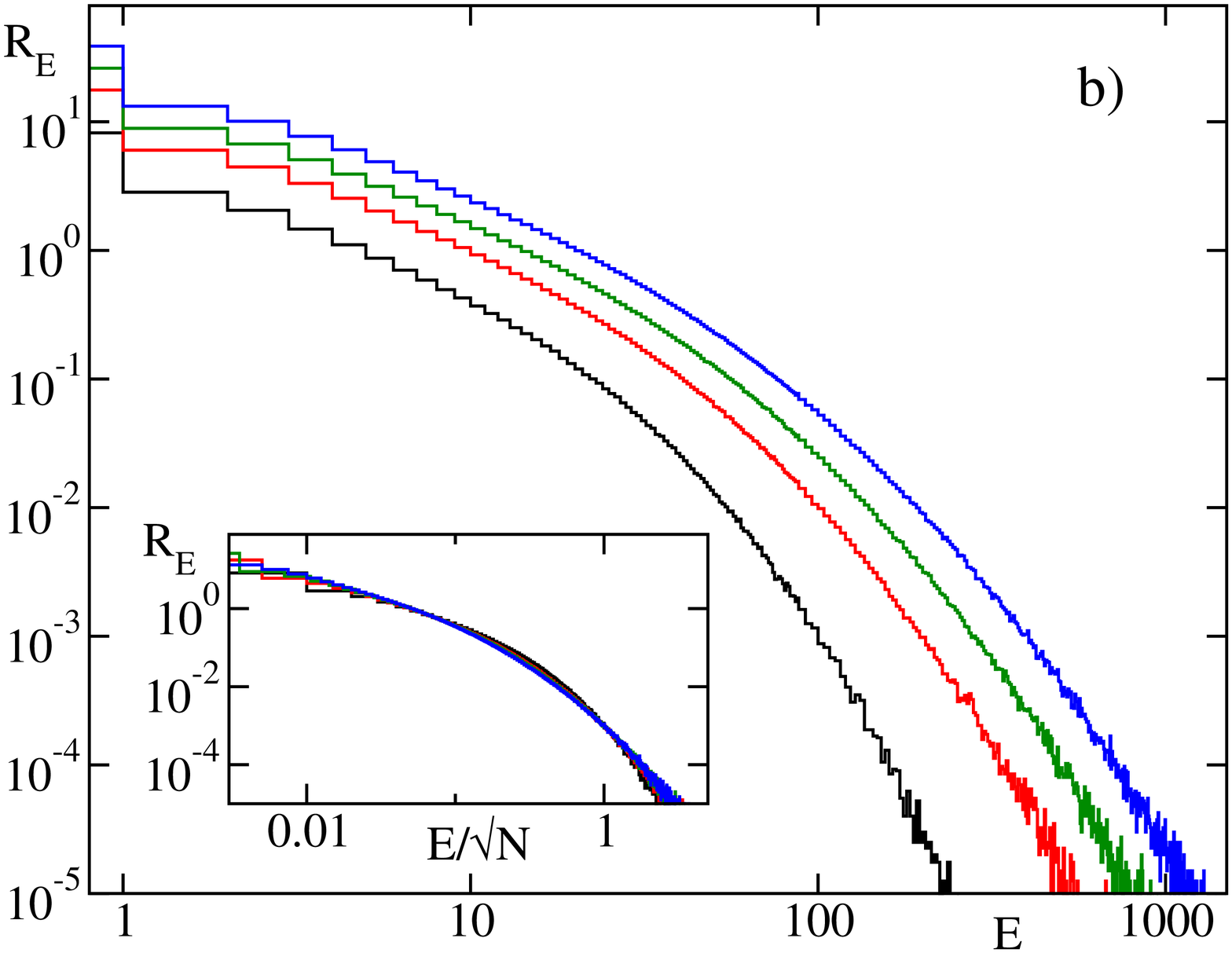}
\end{centering}
\caption{Density per unit of time $R_E$ of simultaneously emitted spikes $E$ for four different system sizes: $N=10,000$ (black), $N=40,000$ (red), $N=160,000$ (green)  and $N=640,000$ (blue) and two different coupling $J=0.2$mV (a) and $0.5$ mV (b).  The time durations of the simulations are set to 100 s. The insets are based on the same data but the abscissas have been rescaled as $E/\sqrt{N}$ to visualise the decreasing effect with the system size. 
\label{fig:simultan}
}
\end{figure}

We first monitored the number of simultaneously emitted spikes $E$ and 
the corresponding density $R_E$ per unit of time; these are shown in Fig.~\ref{fig:simultan}
for different system sizes and two different synaptic coupling values. 
The density $R_E$ obviously increases with the system size. 
As shown in the insets of Fig.~\ref{fig:simultan},
all curves collapse on the same one, once the abscissa has been rescaled as
$E /\sqrt{N}$. This means that the number of simultaneous events $E$ is of order
$\mathcal{O}(\sqrt{N})$, while the total number of emitted spikes is of order $\mathcal{O}(N)$.
Therefore, the strictly simultaneous events become less and less relevant, while 
approaching the thermodynamic limit.

\begin{figure}
\begin{centering}
\includegraphics[width=0.22\textwidth,clip=true]{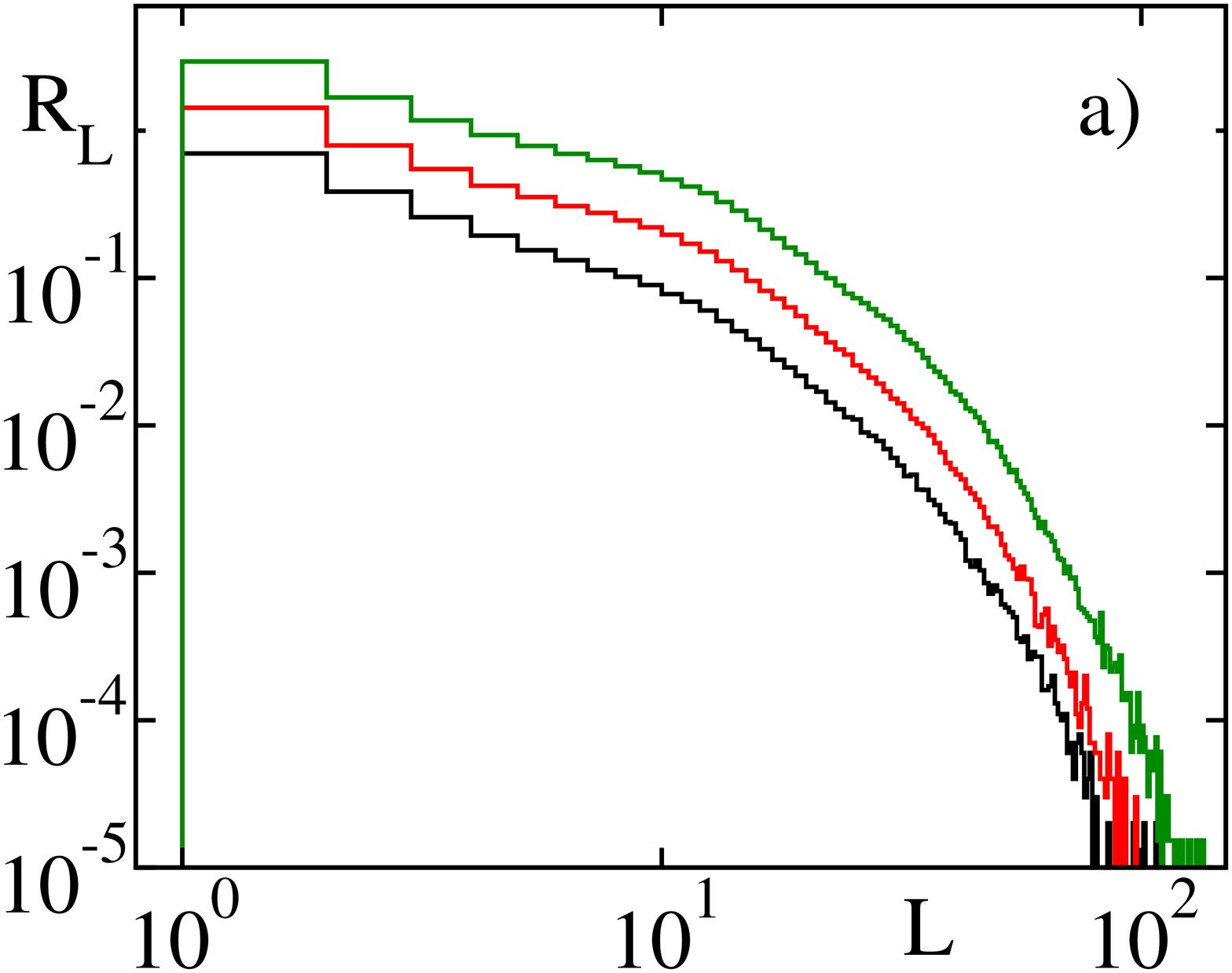}
\includegraphics[width=0.22\textwidth,clip=true]{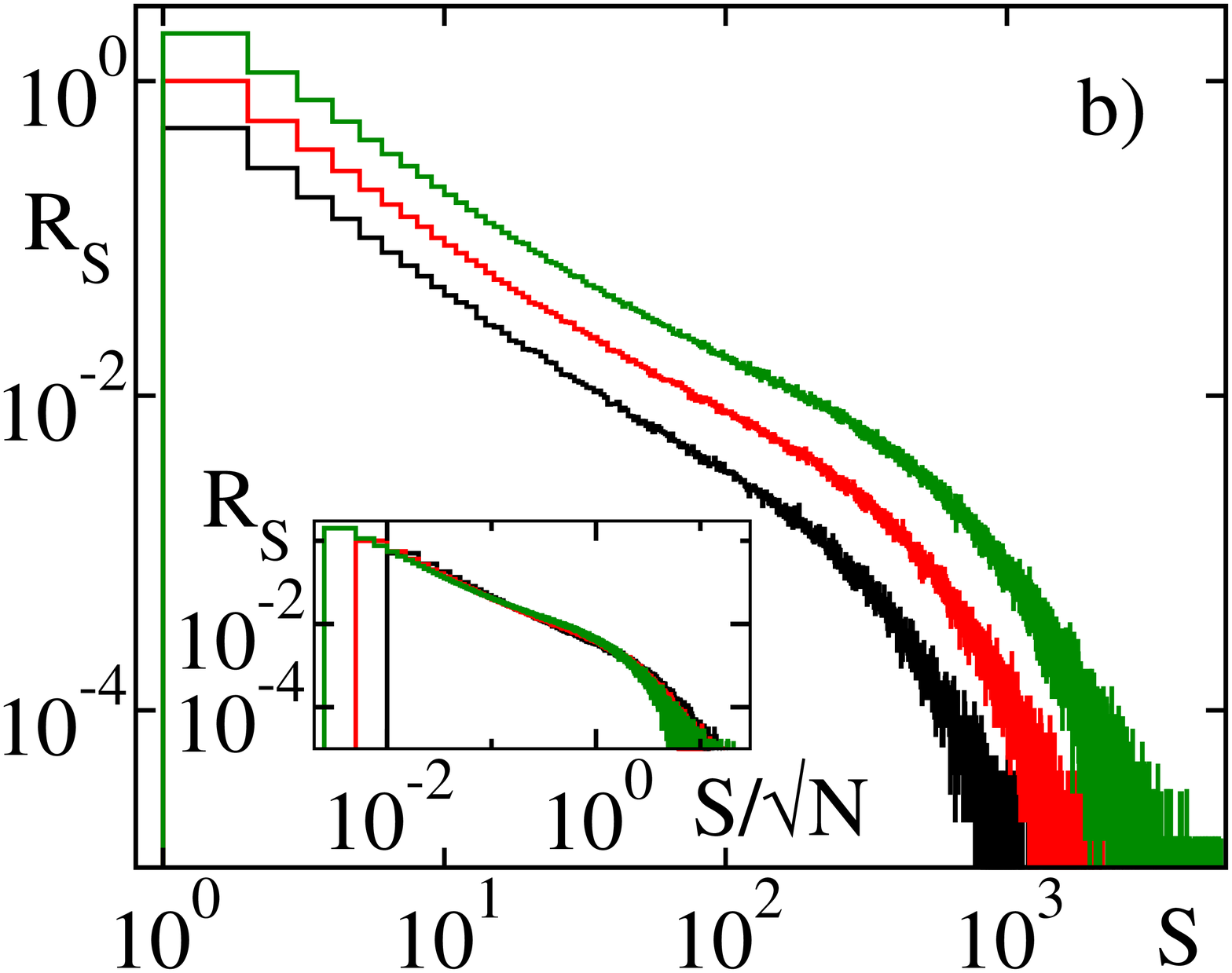}
\end{centering}
\caption{Density per unit of time of the length ($R_L$, a) and size ($R_S$, b) of avalanches for synaptic coupling $J=0.5$ mV. The length of the avalanche $L$ has been expressed in multiple of the time delay.  
Different system sizes are color coded: $N=10000$ (black), $N=40000$ (red) and $N=160000$ (green). 
The time durations of the simulations are fixed to 100 s. The inset of panel (b) is based on the same data as the main one but the abscissa has been rescaled as $S/\sqrt{N}$ to visualise the decreasing effect with the system size.
\label{fig:aval}
}
\end{figure}

The presence of avalanches is a consequence of instantaneous synapses and identical time-delays.
Avalanches arise when a spike triggers a cascade of sub-sequent spikes occurring
at times exactly separated by the time delay. 
We have monitored the time duration (called length $L$) of the avalanches as well as
their consistency (size $S$), corresponding to the total number of spikes
emitted during an avalanche. The densities $R_L$ and $R_S$ per unit of time of the avalanche
length $L$ and size $S$ are reported in Fig.~\ref{fig:aval} for various
system sizes. There is a clear increase of the length and size of the avalanches 
with the system size $N$ and hence (due to the massive coupling) with the in-degree $K=cN$. 
Upon rescaling the $R_S$ densities by $\sqrt{N}$, they collapse onto a same 
curve (that we expect to depend on the coupling strength $J$) as shown
in the inset of Fig.~\ref{fig:aval} (b).
Finally, no simple scaling behavior has been found for the avalanche length $L$.
We tested different scalings assumptions but none of them yielded a convincing
data collapse.
We can nevertheless safely conclude that the length grows even slower than 
the size with $N$. 
Altogether, our results show that the avalanches unavoidably appear also in the exact event driven approach but do not contribute significantly to the network dynamics in the thermodynamic limit.

\section{The Microscopic Dynamics} \label{sec:micro} 

\begin{figure}
\begin{centering}
\includegraphics[width=0.43\textwidth,clip=true]{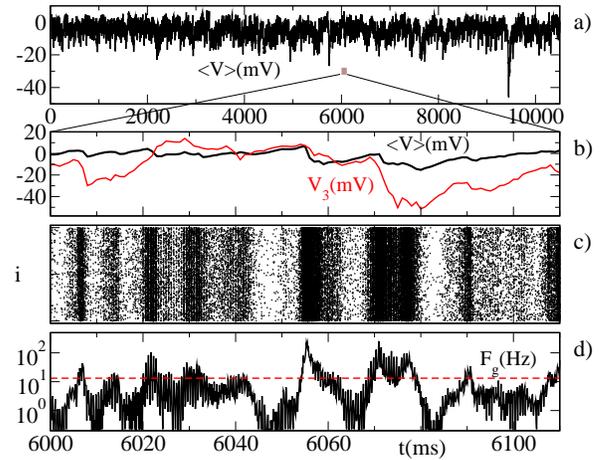}
\end{centering}
\caption{\label{fig:ts_rast}
Characterization of the network dynamics. 
Mean potential $\langle V \rangle$ versus time (a); a zoom in time of the evolutions
is shown in (b) together with the time evolution of the single membrane potential $V_3$ taken from a generic sample neuron identified by the index $3$ (red).
For the same time interval also the raster plot of the whole activity is
reported in (c) as well as a measure of the global spiking activity $F_g$
in (d). The dashed red line in (d) corresponds to the average firing rate, $\overline{\nu}_0 = 13.2$ Hz.
The data refer to $N=40,000$ and $J=0.5$ mV, while the time window employed 
for the measure of $F_g$ is $0.11$ ms.
}
\end{figure}

The CID is a macroscopically observable phenomenon originated by an orchestrated interplay of the 
microscopic oscillators. Before introducing appropriate indicators to characterize the collective phenomena 
(see the following Section~\ref{sec:col}), we first shed light on the microscopic dynamics.
Each oscillator $i$ is characterised by a membrane potential $V_i$ which evolves continuously
in time, but is affected by discrete events associated with the emission times $\{t_{i(n)}\}$ 
of the single spikes and the corresponding arrival times. 
A time trace of the individual membrane potential $V_3$ is shown in Fig.~\ref{fig:ts_rast}(b) 
as a red line together with the mean potential $\left<V\right>(t) = 1/N \sum_i{V_i(t)}$ (black line) 
(here and in the following the symbol $\langle \cdot \rangle$ stands for an ensemble average). 
The membrane potential of a single neuron exhibits significantly larger fluctuations 
than those exhibited by the mean value $\langle V \rangle$ with only a limited correlation among the
two observables (see Fig.~\ref{fig:ts_rast}(b)).
The raster plot in the same time interval is depicted in Fig.~\ref{fig:ts_rast}(c); it clearly 
reveals irregular population bursts, whose degree of synchronization can be appreciated 
by looking at the global firing activity $F_g$, i.e. the number of spikes emitted in a fixed time 
window per neuron (see Fig.~\ref{fig:ts_rast} (d)). This last entity, whose time average corresponds
to the average firing rate, reveals clear irregular oscillations.
 
\begin{figure}
\begin{centering}
\includegraphics[width=0.43\textwidth,clip=true]{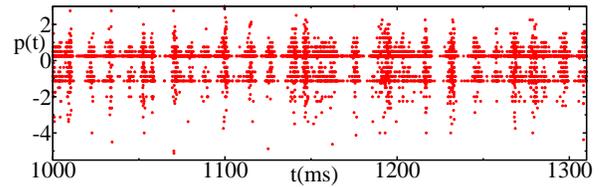}
\end{centering}
\caption{\label{fig:input}
Short time sequence of the post-synaptic inputs $p(t)$ 
received by  a single oscillator for $J=0.5$ mV and $N=40,000$.}
\end{figure}

As explained in Section~\ref{sec:model} and shown in Fig.~\ref{fig:simultan}, simultaneous spike events 
are intrinsic properties of the model and can lead to the simultaneous arrival of a mixture of excitatory 
and inhibitory PSPs. 
Hence, the net result $p(t) = RI_i / \tau$ can be either positive or negative. 
Figure~\ref{fig:input} shows examples of the post-synaptic input $p(t)$ for a synaptic coupling $J=0.5$ mV 
in a system of $N=40,000$ neurons. The most probable input corresponds to a single either excitatory, 
or inhibitory PSP. The other discrete $p(t)$ values are related to all possible combinations of
excitatory and inhibitory PSPs.
 
\begin{figure}
\begin{centering}
\includegraphics[width=0.43\textwidth,clip=true]{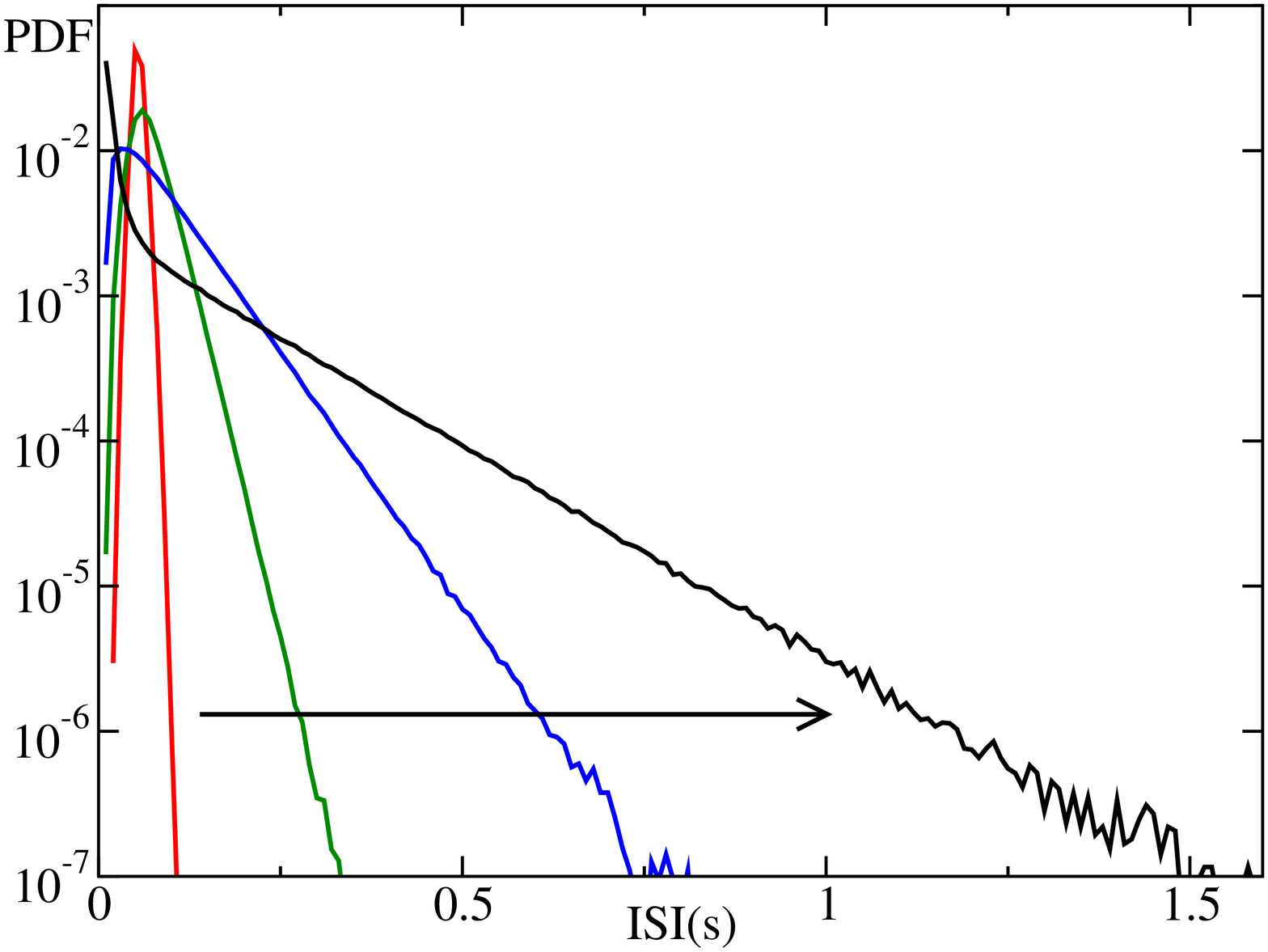}
\end{centering}
\caption{\label{fig:histo}
PDFs of the ISIs for 
a typical neuron for $N=10,000$ and various coupling strengths, namely $J=0.05$ mV (red), $J= 0.1$ mV (blue), $J= 0.2$ mV (magenta) and $J= 0.5$ mV (black). The arrow
denotes the increase of $J$.
}
\end{figure}

Additional information on the variability of the spiking activity is contained in the
probability distribution function (PDF) of the ISIs of a single generic 
neuron. A few PDFs obtained for $N=10,000$ for different coupling strengths are reported 
in Fig.~\ref{fig:histo}.
For sufficiently large coupling, the PDF is composed of an exponential tail characteristic 
of a Poissonian dynamics plus a peak at very short ISI, which is the 
consequence of the occasionally periodic bursting activity of the neurons.
These results clearly indicate that the single neuron dynamics is increasingly 
dominated by fluctuations for large $J$.

There are at least two ways to explore the spectral properties of the single neurons: by looking at the evolution of the membrane potential or by recording the spike events. We have focussed on the latter one, since it allows for a comparison between the input stimulus and the output response, as well as for the analysis of the collective spiking activity (discussed in the next Section). 

\begin{figure}
\begin{centering}
\includegraphics[width=0.43\textwidth,clip=true]{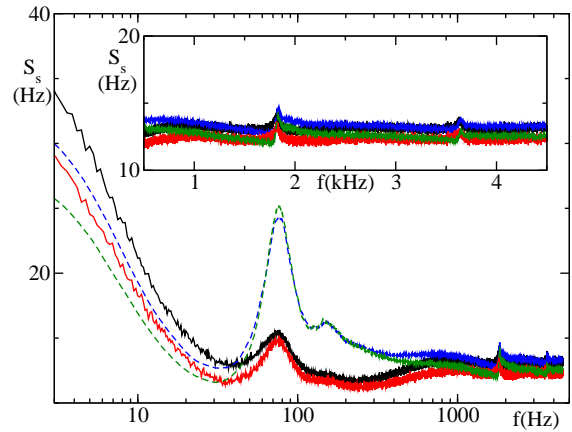}
\end{centering}
\caption{\label{fig:inout_psd}
Spike train spectrum $S_s$ of the input (blue and green dashed lines) and output (black and red solid lines) of a single neuron within an ensemble of $N=40,000$ and $160,000$, respectively, with $J=0.5$ mV. The results has been obtained from time series of $100$ s duration and averaged over $20$ different neurons. 
The inset is the linear representation of the main plot with a focus on high frequencies. 
}
\end{figure}

In order to characterize the output signal we counted
the number of emitted spikes within a fixed time window
(we set it equal to $0.11$ ms). The resulting signal is a series of $0$s interspersed with a few $1$s. The input is instead determined from the 
values of $p(t)$ (reported in Fig.~\ref{fig:input}) coarse-grained over time bins of $0.11$ ms. Furthermore, the power spectrum of the input signal has been rescaled according to the number of excitatory and inhibitory connections and their strength, i.e. by the
factor $cN(b J_e^2 + (b-1) J_i^2)$, to be comparable with the spectrum $S_S(f)$ corresponding to the output time series. 
Figure~\ref{fig:inout_psd} reports the spectra $S_S(f)$ 
associated to the input and to the output signal of a single neuron, 
averaged over $20$ neurons randomly sampled out of $N=40,000$ for the coupling $J=0.5$ mV.
At sufficiently high frequencies ($\gtrapprox 500$ Hz) the input and output spectra almost coincide
for all system sizes and are basically flat and converge towards the average firing rate ${\bar \nu}_0 = 13.2$ Hz, as expected and shown
in the inset of Fig.~\ref{fig:inout_psd}.

At lower frequencies, and especially for $50 \le f \le 500$ Hz, the $S_S(f)$ spectra of input and 
output differ from one another (the differences persist for $N=160,000$, where they have
reached an asymptotic shape  -- data not shown):
in particular the input spectra exhibit a clear peak at $\simeq 75$ Hz,
while the output ones reveal just a shoulder.
In the presence of an asynchronous regime, input and output spectra should coincide 
(except for a scaling factor). In fact, in a series of recent papers, a recursive method 
was developed to generate asymptotic spectra, exactly by imposing a perfect correspondence
between input and output~\cite{dummer2014,wieland}.
The clear difference shown by our numerical results (Fig.~\ref{fig:inout_psd}) provides a first
indication of a collective dynamics or, otherwise stated, of nontrivial correlations
among the different neurons.

This will be extensively elucidated in the next Section, where we computed various
indicators, including the power spectrum of the overall activity, whose spectrum
is not too different from that of the input (see Fig.~\ref{fig:power}).

\section{Collective dynamics} \label{sec:col}  

In this section we discuss the collective dynamics which emerges from the
correlations in the microscopic activity of the single neurons.
A qualitative evidence is already noticeable in the structure of a typical raster plot,
which consists in an irregular alternation of regions of different density
(see Fig.~\ref{fig:ts_rast}(c)).
The evolution of $\langle V \rangle$ provides a more accurate representation.
The fluctuations of $\left<V\right>$ are in fact smaller than those of the individual membrane 
potential $V_i(t)$ (see the red line in Fig.~\ref{fig:ts_rast}(b)), but nevertheless 
definitely appreciable.

On a quantitative level, it is convenient to introduce the synchronisation measure $\rho$ 
\begin{equation}
\rho^2 \equiv \frac{\overline{\langle V\rangle^2}-\overline{\langle V\rangle}^2}
    {\langle \overline{V^2}-\overline{V}^2\rangle} \; ,
\end{equation} 
where the overbar denotes a time average. 
Perfectly synchronised neurons behave in exactly the same way, so that the numerator and the denominator 
are equal to one another and $\rho=1$. If instead, they are statistically independent, 
$\rho \approx 1/\sqrt{N}$.

\begin{figure}
\begin{centering}
\includegraphics[width=0.43\textwidth,clip=true]{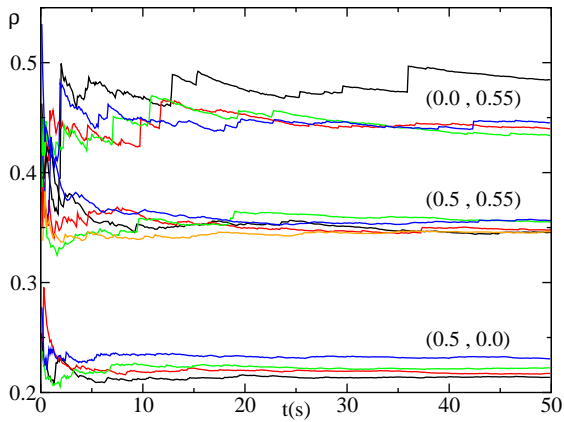}
\end{centering}
\caption{\label{fig:NOpar}
Running average of the order parameter $\rho$ for increasing integration time $t$, obtained after discarding a transient of at least $5$ s for $J=0.5$ mV for different system sizes and parameter settings labeled by the tuple ($\tau_r$ , $\tau_d$). The bunch of curves reported in the middle of the figure corresponds to the standard setup with delay $\tau_d = 0.55$ ms and refractoriness $\tau_r= 0.5$ ms. The upper family refers to a setup with standard delay $\tau_d=0.55$ ms but without refractoriness, i.e. $\tau_r=0$ms. The lower collection of lines corresponds to no delay $\tau_d=0.0$ ms but with refractoriness $\tau_r=0.5$ ms. The system sizes are color coded in ascending order: black, red, green, blue and orange for $N=10,000$, $20,000$, $40,000$, $80,000$ and $160,000$, respectively.}
\end{figure}

The progression of the running average of $\rho$ can be appreciated in Fig.~\ref{fig:NOpar} for $J=0.5$ mV, see the middle bunch of trajectories, labeled by (0.5 , 0.55). The order parameter approaches $\rho \approx 0.35$, irrespective
of the networks size, indicating that CID is not a finite-size effect, but survives in the
thermodynamic limit as defined in Ref.~\cite{noi}.
The jumps observed at early times of the running average are caused by sudden drops of the mean potential $\left<V\right>$. One of the strong drops is shown in Fig.~\ref{fig:ts_rast}(a) 
around $t=9,500$ ms. These rare events appear randomly at all times, but their effect on the
cumulative average obviously decreases as time progresses.

The coefficient of variation $C_v$ is another measure of irregularity of the dynamics, based on the 
fluctuations of the ISI, rather than of the membrane potential. More precisely, we calculate
\[
\left<C_v\right> = \langle \frac{\sigma_S}{\tau_S} \rangle
\]
where $\sigma_S$ is the standard deviation of the single-oscillator ISI, while $\tau_S$ is the
corresponding mean ISI.

\begin{figure}
\begin{centering}
\includegraphics[width=0.43\textwidth,clip=true]{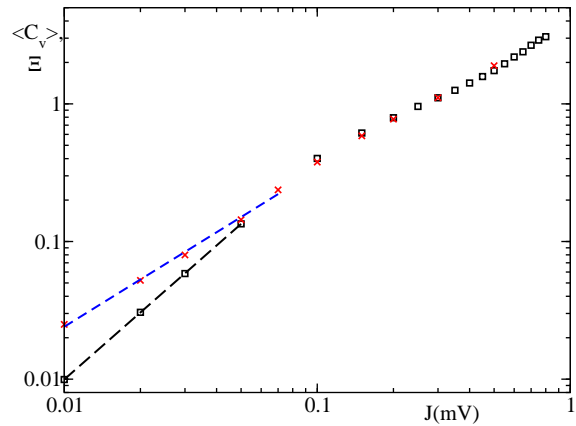}
\end{centering}
\caption{\label{fig:cvISI}
The mean coefficient of variation of the interspike interval $\left<C_v\right>$. The black squares refer to $\left<C_v\right>$ for $N=10,000$ and the red crosses show the square root of the rescaled diffusion coefficient $\Xi$. Power law fits $J^{\alpha}$ are denoted with dashed lines. The black dashed line power law fit matches the lower part of $\left<C_v\right>$ with exponent $\alpha = 1.62$ very well whereas the blue dashed lines refers to a power law fit with exponent $\alpha = 1.14$.}
\end{figure}

For $J=0$ the single-neuron activity is strictly periodic and thus $C_v$ is equal to zero.
We expect it to increase when the coupling strength $J$ is switched on. 
For small $J$ we can indeed appreciate a power-law growth, $\left<C_v\right> \approx J^\alpha$
(see the black squares in Fig.~\ref{fig:cvISI}) with a value of the rate $\alpha$ close to $1.6$.
The growth of $\langle C_v \rangle$ continues for stronger coupling strenghts, becoming larger than 1,
the value expected for a Poisson statistics.
For sufficiently large coupling, we observe bursting dynamics of the neurons,
corresponding to $\langle C_v \rangle > 1$.  A similar behavior
of $\langle C_v \rangle$ with the coupling strength has been reported for inhibitory
sparse networks of LIF in the absence of delay but for sufficiently slow 
synaptic decays \cite{david2017}.
Interestingly, the $\langle C_v \rangle$ values are substantially independent of
the system size (we have tested values of $N$ up to $640,000$).

The coefficient of variation measures the amplitude of the fluctuations, but it is insensitive to
temporal correlations: $\langle C_v \rangle$ is strictly larger than zero already for a very regular
sequence of ISIs such as a periodic alternation of two values $t_1$ and $t_2$, with $t_1\ne t_2$.
In order to have a more accurate indicator, we have computed the following diffusion coefficient.
Let $T_n$ denote the time of $n$-th spike emitted by a given neuron, so that $T_{n}-T_{n-1}$ is
the $n$-th ISI.
Let, then
\begin{equation}
D_S =  \lim_{n\to \infty} \frac{(T_n-n \tau_S)^2}{n}
\end{equation}
be the diffusive coefficient of the process $T_n$. We finally define
\begin{equation}
\Xi \equiv  \frac{\sqrt{D_S}}{\tau_S}  \; ,
\end{equation}
$\Xi$ is plotted in Fig.~\ref{fig:cvISI} (see the red crosses). Above $J=0.1$,
it basically coincides with $C_v$, indicating that $T_n$ is essentially
a renewal process. For $J\le 0.1$, $\Xi$ decreases more slowly (with
a rate close to 1.14) than $C_v$. This is clearly due to the increasingly
periodic character of the dynamics.
The overall scenario is reminiscent of the phase transition discussed by
Ostojic in \cite{ostojic}.

\begin{figure}
\begin{centering}
\includegraphics[width=0.43\textwidth,clip=true]{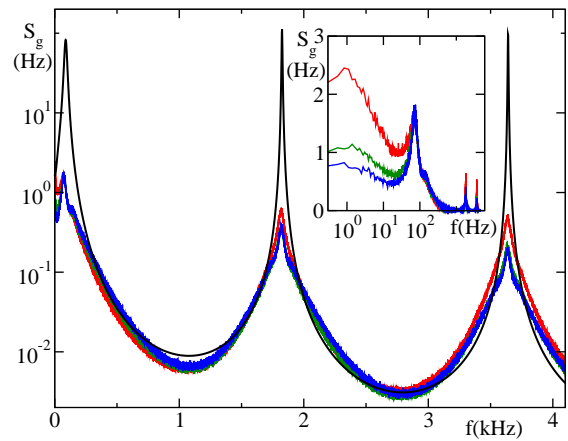}
\end{centering}
\caption{\label{fig:power}
Power spectra $S_g$ of the global activity for $J=0.5$ mV. The spectra 
reported for different system sizes $N=10,000$, $N=40,000$ and $N=160,000$ are shown in red, green and blue, respectively. The black line represents the results from the Brunel's perturbative theory (see Appx.~\ref{app:pert}) and it is reported only in the main plot and not in the inset.}
\end{figure}

The power spectrum of the global activity $S_g$ sheds light on collective phenomena from yet a different 
perspective. Analogously to the single oscillators, the spike times have been converted into a 
single time series counting the number of spikes emitted in each time bin. We chose the same time bin of $0.11$ ms as in the previous cases. 
An example of the global field $F_g$ is included in Fig.~\ref{fig:ts_rast} in the bottom panel (d); 
it clearly shows an irregular behavior. 
The power spectrum of the global activity has been divided by $N^2$ to allow for a meaningful comparison amongst different system sizes and with the single-neuron spectrum  (Fig.~\ref{fig:inout_psd}).
The spectra obtained for different system sizes are plotted in Fig.~\ref{fig:power}. For $f>40$ Hz, they collapse onto one another, suggesting that the dynamics remains irregular 
in the thermodynamic limit, i.e. that the fluctuations exhibited by the collective variables are not finite-size effects. It should be stressed that in the absence of collective effects, i.e. for asynchronous states, 
the spectrum of the global activity would be proportional to $N$ rather than to $N^2$.

Below $40$Hz, the spectral amplitude decreases with the system size, suggesting 
that the zero-frequency peak eventually disappears (at least for $J=0.5$).
Altogether, the spectral power is mostly concentrated in two frequency ranges: (i) a broad peak around 
$f \approx 75$ Hz, which corresponds to the peak observed in the single
neuron spectra $S_S$ and is presumably related to a time scale of the order of the
membrane time constant $\tau = 20$ msec  and (ii) a peak around 
$f \approx 1818$ Hz (and its multiples), which corresponds to the inverse of the delay. 
A comparison with the spectrum of the single neuron activity (see Fig.~\ref{fig:inout_psd}),
reveals that the latter one is characterized by a much stronger high-frequency component
(of white-noise type) and weaker peaks in correspondence of the inverse delay time.

We have finally implemented the perturbative approach developed by Brunel~\cite{brunel2000},
based on the assumption of a sparse coupling. The idea basically consists in solving a self-consistent
noisy Fokker-Planck equation for the probability density $P(v,t)$ of the membrane potential $v$,
\begin{equation}
\tau \frac{\partial P}{\partial t} = \frac{\partial}{\partial v} [(v-\mu-\mu_{e})P]
+ \frac{\sigma^2}{2} \frac{\partial^2P}{\partial v^2}
+ \sigma_0 \sqrt{c \tau }   \frac{\partial P}{\partial v}\zeta(t) \; .
\label{eq:FP}
\end{equation}
The first term in the r.h.s. is nothing but the deterministic current defined in Eqs.~(\ref{eq:LIF},\ref{eq:general}),
with $\mu_e=RI_0$. The second, diffusive contribution, accounts for the unavoidable statistical fluctuations
of the input signal arising from the coupling with the other neurons and its amplitude is estimated under the
assumption of being a Poisson process. Finally, the last term is a common noise due to the fact that different neurons
partially share the same input signals, whenever they share the same afferent neurons.
More precisely, the drift $\mu$ is defined as 
\begin{equation}
\mu = -\sqrt{c}\tau (1-b)g_1 \mathcal{J}\nu(t-\tau_d)
\label{eq:mu}
\end{equation}
where $\nu(t)$ is the instantaneous firing rate, while an expression for the diffusion
coefficient $\sigma^2$ can be obtained by assuming that the spike train follows a
Poisson statistics,
\begin{equation}
\sigma^2 =   \frac{b \tau \mathcal{J}^2}{1-b}\nu(t-\tau_d) \; .
\label{eq:sigma}
\end{equation}
Finally, $\sigma_0$ is the value of $\sigma$ corresponding to the stationary value of the firing rate in the asynchronous regime $\nu_0$ and $\mathcal{J} = J\sqrt{K}$ (see the appendix for a more precise definition).
The power spectrum $\hat n(\omega)^2$ of the neural activity can be determined by solving
perturbatively the Fokker-Planck equation. The technical details are presented 
in Appendix~\ref{app:pert}: we practically follow the method introduced in \cite{brunel2000}, the
main difference being the numerical strategy adopted to determine the spectrum.
The resulting curve is shown in Fig.~\ref{fig:power} (black line) after converting the angular frequency $\omega$ into the frequency $f$. 
The perturbative approach qualitatively reproduces the shape of the spectrum, including the position
of the peaks. On the other hand, the height of the peaks strongly deviates from the numerical 
simulations. For large coupling, the agreement worsens and the perturbative approach fails even in reproducing qualitatively the spectra at low frequencies, as shown in \cite{noi}.

\begin{figure}
\begin{centering}
\includegraphics[width=0.43\textwidth,clip=true]{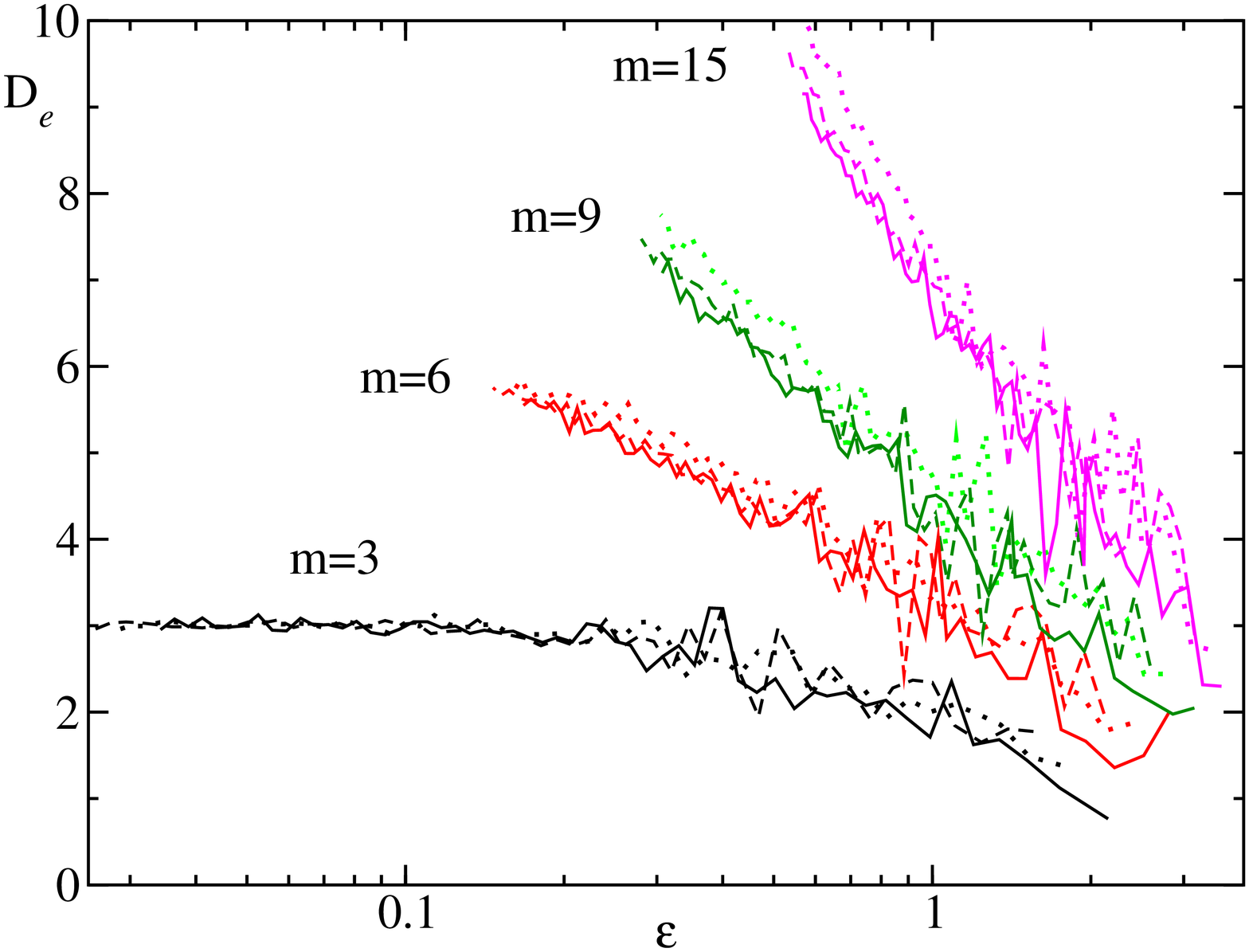}
\end{centering}
\caption{\label{fig:frac}
Effective dimension $D_e$ as a function of the resolution for  $J=0.2$ mV and
different system sizes: $N=10,000$ (dotted), $40,000$ (dashed), and $160,000$ (solid). The different groups of curves correspond to different embedding dimensions $m$.}
\end{figure}

Finally, we have studied the neural activity by implementing nonlinear-dynamics tools,
to determine the (effective) fractal dimension $D_e$ of the
mean potential $\left<V\right>(t)$.
Given the sequence of values $\left<V\right>(t_n)$, obtained by sampling the original signal every
$\Delta t=1$ ms over $500$ s (i.e. resulting in $500,000$ data points), this is embedded into a space of dimension $m$, 
by building vectors of the type $[\left<V\right>(t_n), \left<V\right>(t_{n+1}), \ldots, \left<V\right>(t_{n+m-1})]$.
The fractal dimension has then been estimated by using a variant of the 
nearest-neighbour method recently proposed in \cite{ullner2016}.
In particular, $N_r$ reference points are randomly selected 
($N_r = 10^5$ in our case), then each reference point is compared with other $n$ randomly selected points (up to the number of data points available) determining the distance $\varepsilon_m(k,n)$ of the $k$-th neighbour for different values of $m$ and $k$. The distance is herein estimated using the maximum norm. An  established theory~\cite{badii1985}, implies that for large $n$,
\[
-\frac{\ln n}{\langle \ln \varepsilon_m(k,n)\rangle} \approx D_e \; ,
\]
where the angular brackets denote the average over the reference points, while 
$D_e$ is the information dimension. 
In practice, the logarithmic derivative of $\varepsilon$ varies with $n$ before
eventually converging to its asymptotic value.
Accordingly, it can be interpreted as an effective, resolution-dependent dimension,
which is, in fact, independent of the order $k$ of the neighbour considered in
the simulations.
In practice, given $\varepsilon$, the inverse of the logarithmic derivative is first
determined and then plotted versus the resolution $\varepsilon$.
The results are reported in Fig.~\ref{fig:frac}. They show that, independently  of
the network size, the effective dimension increases upon decreasing the resolution.
The stochastic-like nature of the dynamics is further confirmed.

\section{Robustness} \label{sec:robust} 

In this section we investigate the robustness of CID, by testing its properties when some of the 
model parameters are modified, notably refractoriness, delay, 
connectivity and finally after introducing an external noise which acts independently on each
neuron. 

We start by showing the dependence of the average firing rate $\overline{\nu}_0$
and of the coefficient of variation $\langle C_v\rangle$ on the system size in a setup where
either delay or refractoriness is missing. From the data reported in the Table~\ref{tab:robust}, 
which refer to $J=0.5$mV, we observe that, strange enough, the firing rate slows down, 
when the delay is removed. This is because the absence of delay induces a more homogeneous 
firing activity, which, in turn, is more dominated by the inhibitory neurons due to the weak unbalance. This interpretation is
confirmed by the lower degree of synchronization that can be appreciated by looking at the
(0.5,0.0) curves in Figure~\ref{fig:NOpar}, which refer to different network sizes.
Although $\rho$ decreases, CID is still present in the absence of delay. In fact $\rho$ is substantially
independent of $N$ (it actually even slowly increases). Additional studies of the spectral properties
confirm that the collective dynamics is irregular (data not shown).
This is at variance with the setup studied in~\cite{luccioli2010}, a heterogeneous ensemble of fully coupled, inhibitory, LIF neurons. In that context, CID disappears as soon as the delay vanishes. It is still
to be understood whether the qualitative difference is due to the heterogeneity  (dispersion in the bare firing rates of the single neurons). 

Refractoriness is less relevant. From the data in Table~\ref{tab:robust}, we see that its absence
does neither significantly modify the firing rate (which naturally increases by a small amount),
nor the degree of irregularity of the single neuron.
Even though $\langle C_v \rangle$ slowly decreases upon increasing $N$, it remains substantially 
larger than 1, the expected value for a Poisson statistics. As for the collective dynamics, 
we notice in Fig.~\ref{fig:NOpar} (see the curves labeled (0.0,0.55)) that synchronization increases upon removing refractoriness. The convergence is slower than in
the previous case: this is because of the presence of several sudden burst of synchronizations
(see the upward jumps exhibited by $\rho(t)$), which require longer time scales for them to
be suitably averaged out.



\begin{center}
\begin{table}
\caption{\label{tab:robust} 
Mean firing rates $\overline{\nu}_0$ and mean coefficients of variation of the ISI $\left<C_v\right>$ in absence of delay 
($\tau_d=0$) or no refractoriness ($\tau_r=0$) for different system sizes $N$, for $J = 0.5$ mV. The last two column reference to the standard setup with delay $\tau_d=0.55$ ms and with refractoriness $\tau_r=0.5$ ms.
}
 \begin{tabular}{| C {1.0cm} || C {1.1cm} | C {0.9cm} | C {1.1cm} | C {0.9cm} | C {1.1cm} | C {0.9cm} | } 
 \hline
\multirow{2}{*}{$N$} & \multicolumn{2}{|c|}{$\tau_d=0$} & \multicolumn{2}{|c|}{$\tau_r=0$} & \multicolumn{2}{|c|}{standard}  \\
 \cline{2-7}
  &  $\overline{\nu}_0$ $[Hz]$ & $\left<C_v\right>$  & $\overline{\nu}_0$ $[Hz]$ & $\left< C_v \right>$ & $\overline{\nu}_0$ $[Hz]$ & $\left< C_v \right>$   \\ 
 \hline
 10,000 & 13.8 & 1.68 & 15.9 & 1.80 & 15.3 & 1.75\\ 
 \hline
 20,000 & 13.2 & 1.63 & 14.3 & 1.67 & 14.3 & 1.67\\ 
 \hline
 40,000 & 12.4 & 1.58 & 13.4 & 1.60 & 13.2 & 1.59\\ 
 \hline
 80,000 & 11.9 & 1.54 & 13.0 & 1.55 & 12.8 & 1.55\\ 
 \hline
\end{tabular}
\end{table}
\end{center}

\begin{figure}
\begin{centering}
\includegraphics[width=0.43\textwidth,clip=true]{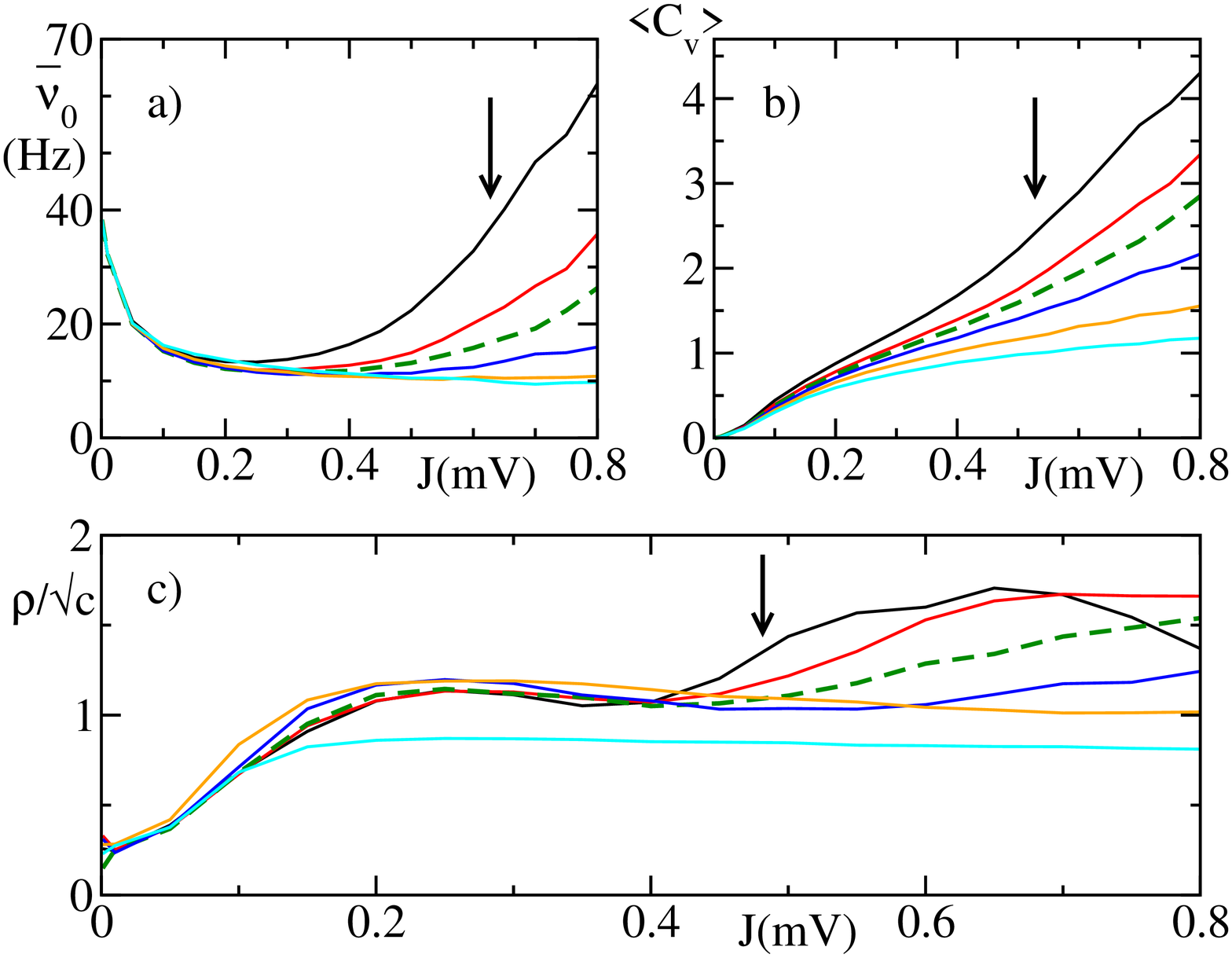}
\end{centering}
\caption{\label{fig:connect}
The mean firing rate $\overline{\nu}_0$, the mean coefficient of variation in the ISI $\left<C_v\right>$ and the synchronisation measure $\rho$ versus the coupling $J$ for different network connectivities $c$. The network connectivity $c$ follows in ascending order the direction of the arrow according $c=0.01$, $0.05$, $0.1$, $0.2$, $0.4$ and $0.6$. The usually used $c=0.1$ has been singled out by dashed lines. The system size is $N=40,000$ for all simulations.}
\end{figure}

After having verified that neither delay nor refractoriness are necessary ingredient for CID to be
observed, we now explore the role of the connectivity $c$.
In Fig.~\ref{fig:connect} we plot three key parameters (the firing rate $\overline{\nu}_0$, $C_v$ and $\rho$)
as a function of the coupling strength $J$ for different $c$ values.
In panel (a), we see that the firing rate is almost independent of $c$ in the small coupling limit,
while it progressively decreases upon increasing $c$ in the strong coupling regime. This is due to the
fact that a strong connectivity reduces the fluctuations which are known to be responsible
for the larger $\overline{\nu}_0$ observed for strong coupling~ \cite{Mastrogiuseppe2017ploscb}.
The progressive regularization of the neural activity is confirmed in panel (b), where we see
that $\left<C_v\right>$ decreases upon increasing the connectivity.

Quite interesting is the dependence of the order parameter $\rho$ on $c$. The clean data 
collapse for $J < 0.4$, indicates that up to a $30\%$ connectivity, $\rho$ scales as 
$\sqrt{c}$, in agreement with the perturbative theory developed in~\cite{brunel2000} and
briefly recalled in Appx.~\ref{app:pert}, which predicts a power spectrum proportional to $c$.
The strong coupling regime ($J>0.4$) seems to be characterized by different scaling properties,
but additional simulations for different network sizes are required to put the statement on a
more firm basis.

\begin{figure}
\begin{centering}
\includegraphics[width=0.43\textwidth,clip=true]{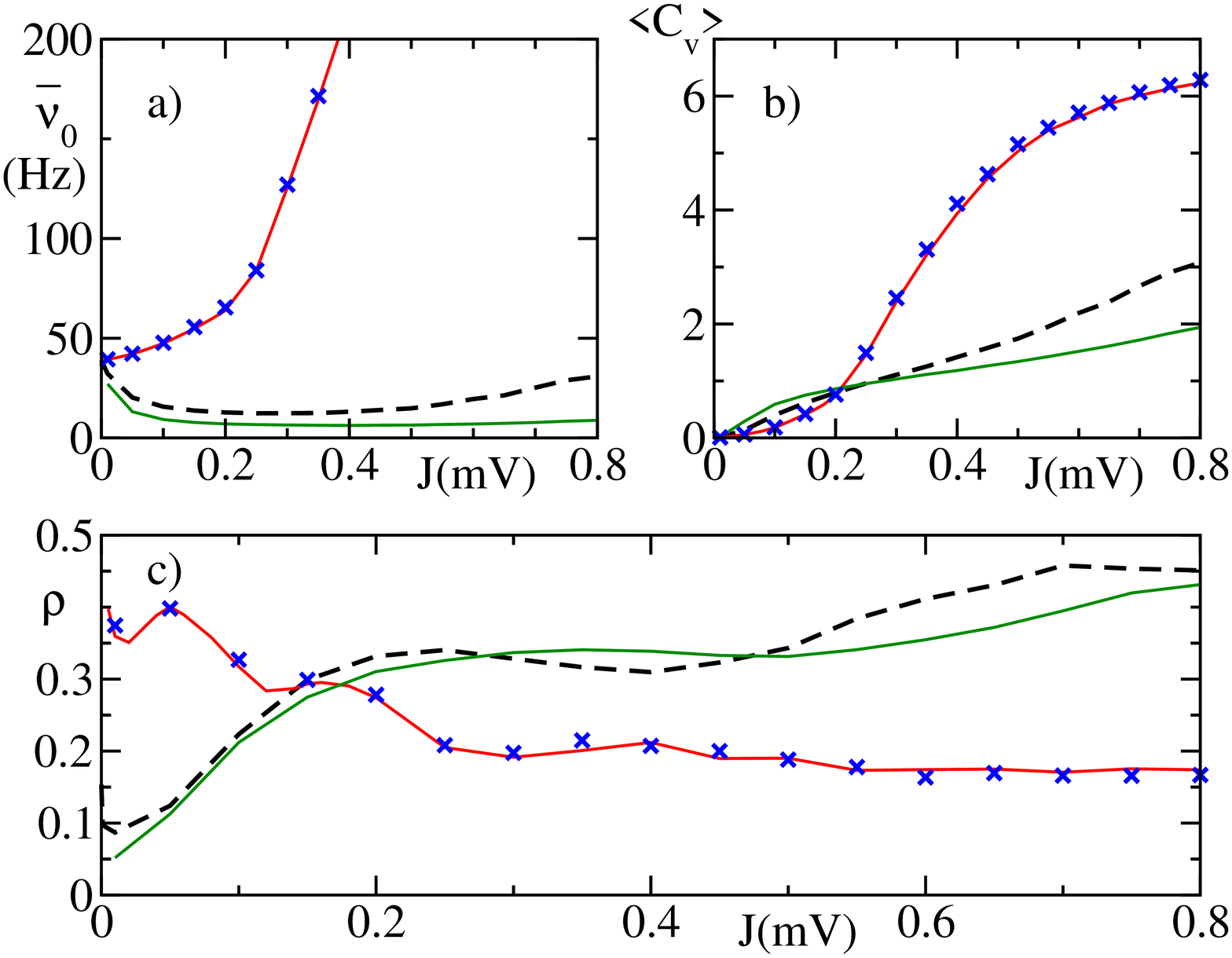}
\end{centering}
\caption{\label{fig:balance}
The mean firing rate $\overline{\nu}_0$, the mean coefficient of variation of the ISI $\left<C_v\right>$ and the synchronisation measure $\rho$ versus the coupling strength $J$ for different balance factors $g$
for a system size $N=10,000$. 
The dashed black line reference to the slight unbalance $g=5$ used throughout the paper, whereas the solid green line shows the situation for stronger unbalance ($g=6$) and the solid red line corresponds to the perfect balanced setup ($g=4$). 
The blue crosses refer to a system size $N=40,000$ for a perfectly balanced situation, i.e. $g=4$.}
\end{figure}

So far, our simulations have been performed for a slight prevalence of the inhibitory activity.
In fact (for $N=10^4$) the ratio between the two coupling strengths is $g \equiv J_i/J_e = 5$, 
to be compared with a $1:4$ ratio of the two corresponding populations.
In order to investigate the role of the degree of unbalance, we have studied two additional $g$-values,
$g=4$, and 5, which, respectively, correspond to a perfect balance and a stronger prevalence of inhibition.
The results are presented in Fig.~\ref{fig:balance}. The most important point is that CID is present
for both parameter values, confirming the robustness of this phase.
On a more quantitative level, unsurprisingly, the perfectly balanced state is characterized by a much
stronger firing activity.

\begin{figure}
\begin{centering}
\includegraphics[width=0.43\textwidth,clip=true]{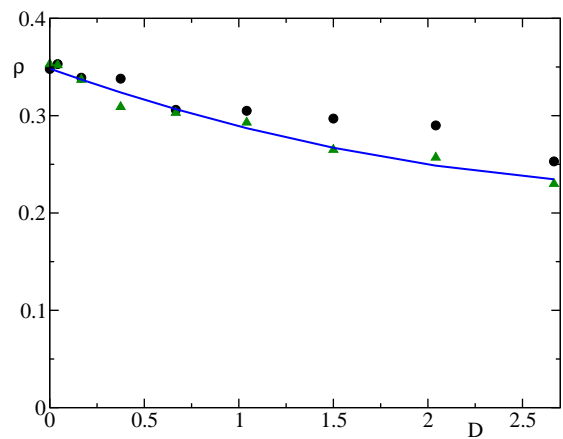}
\end{centering}
\caption{\label{fig:noise}
Order parameter $\rho$ for $J=0.5$ mV and different noise levels $D$ for $N=10,000$ (black circuits) and $40,000$ (green triangles). The solid line is a quadratic fit.}
\end{figure}

Finally, we analyse the role of noise, by adding iid white noise terms $\xi(t)$ to the
single neuron dynamics (Eq.~(\ref{eq:LIF})), such that $\langle \xi(t+\tau)\xi(t)\rangle = 2D \delta(\tau)$.
The dependence of $\rho$ on $D$ is reported in Fig.~\ref{fig:noise}, for two different network sizes.
The noise tends obviously to decrease the strength of the collective dynamics, without, however,
killing it. In fact, CID survives even for moderately strong noise amplitudes, as
it is appreciated by seeing that $\rho$ does not vary significantly upon increasing $N$.

Altogether, CID is a very robust property, which survives even when noise is added, the connectivity is decreased, the balance is changed, the delay or refractoriness removed from the model equations.


\section{Conclusions and open problems}\label{sec:conclusion} 

In this paper we have presented an extensive analysis of the collective dynamics
emerging in a quasi-balanced network of LIF neurons.
The irregularity of the collective dynamics is testified not only by the power spectra 
of the neural activity but also by a fractal-dimension analysis.
The detailed simulations performed for different parameter values confirm
that irregular dynamics is very ubiquitous.
Several questions are, however, still open. Here we list the main ones.

(A) To what extent is this irregular dynamics related to the similar regime observed in
globally coupled, heterogeneous neurons \cite{luccioli2010,ullner2016}? In those setups, which are
reminiscent of the Kuramoto model, the heterogeneity seems to be a crucial ingredient, since CID 
disappears when the diversity among the neurons is removed. Here, it seems that the collective,
stochastic-like dynamics is the result of a microscopic pseudo-chaotic evolution, which
percolates up to macroscopic scales, as a consequence of the quasi-balanced regime. 
Whether this is really the correct explanation it is however still unclear.

(B) All the models so far explored assume $\delta$-pulses, but this is obviously an approximation.
The limit of infinitely narrow PSPs is singular, as shown, for instance, while investigating the stability 
of the splay state~\cite{olmi2012}. Furthermore, we have seen that the presence of strictly $\delta$-like pulses 
induces unavoidable synchronous events whose treatment requires additional ad-hoc hypotheses. 
It will therefore, be instructive to explore networks characterized
by PSPs of finite duration, e.g by considering exponential or $\alpha$-pulses.

(C) The numerical analysis has revealed that collective dynamics arises also for a very small
coupling strength. The weak-coupling limit is typically amenable to a perturbative treatment.
Accordingly, it is plausible that a model of Kuramoto-Daido phase-oscillators might be able
to reproduce a similar regime and, at the same time, allow for an analytical treatment.

(D) The only limit where the irregular collective dynamics vanishes is that of a sparse network, where
$K/N \to 0$ for $N \to \infty$. However, this statement refers to random Erd\"os-R\'enyi-type 
networks. It would be interesting to explore different more elaborated network structures as
well as the role of heterogeneity.

(E) Qualitatively speaking, it looks like some differences exist between the weak and strong 
coupling regime. In the former case, the single neuron spiking activity is strongly correlated 
(being far from a renewal process) as shown in Fig.~\ref{fig:cvISI} and the strength of the collective dynamics 
is reproduced as expected by the perturbative theory (see the nice overlap among the curves reported 
in panel (c) of Fig.~\ref{fig:connect}). In the latter case, the neuronal activity is very well approximated by a
renewal process and, at the same time, the perturbative theory seems to fail already for
a 1\% connectivity. 
These differences suggest that at small coupling the neuronal dynamics is mean driven,
i.e. is dominated by the mean value of the DC currents, while at large coupling it
is fluctuation driven, i.e. the neurons are in proximity or below
the threshold and the firings are triggered by fluctuations of the input currents.
Similar transitions from mean to fluctuation driven dynamics
has been recently reported in sparse inhibitory heterogeneous networks made of LIF neurons
in \cite{david2017} and composed of realistic models of striatal medium spiny neurons in \cite{ponzi2013}.
An additional finite-size analysis is necessary to test whether this is a true transition that persists
in the thermodynamic limit, as claimed by Ostojic~\cite{ostojic} for strongly diluted networks.

(F) In all of our simulations, excitatory and inhibitory neurons have been assumed to be equal to one
another. This implies that the same combination of excitatory and inhibitory fields is automatically
consistent with the evolution of both types of neurons.
This strong limitation should be lifted before drawing yet more general conclusions about the ubiquity
of collective irregular dynamics.

\section*{Acknowledgments}

The authors acknowledge N. Brunel, F. Farkhooi, G. Mato, S. Ostoijc, A. Roxin, and M. di Volo for useful discussions. One of us (AT) has been supported by the French government under the Excellence Initiative I-Site Paris Seine (No ANR-16-IDEX-008) and under the
Labex MME-DII (No ANR-11-LBX-0023-01). The work has been
mainly realized at the Max Planck Institute for the Physics of Complex Systems (Dresden, Germany) during the Advanced Study Group 2016/17 ``From Microscopic to Collective Dynamics in Neural Circuits”.

\appendix
\section{Perturbative approach} \label{app:pert}

Starting from the Fokker-Planck equation~(\ref{eq:FP}), the
firing rate $\nu(t)$ can be also expressed in terms of the probability current at $v=v_{\theta}$, i.e.
\[
\frac{\partial P}{\partial v}(v_\theta,t) = -\frac{2\nu(t)\tau}{\sigma^2(t)}
\]
where $P(v_\theta,t) =0$.
Additionally, the probability density must be continuous at the reset potential $v_r$,
where there is an additional current due to the neurons ending their refractory period
\[
\frac{\partial P}{\partial v}(v_r^+) -
\frac{\partial P}{\partial v}(v_r^-) = -\frac{2\nu(t-\tau_r)\tau}{\sigma^2(t)}
\]
The list of boundary conditions is completed, by including
\[
\lim_{v\to-\infty} P(v,t) = 0
\]
and the normalization
\[
\int_{-\infty}^{v_{\theta}} dv P(v,t) + p_r(t) = 1
\]
where 
\[
p_r(t) = \int_{t-\tau_r}^t du \nu(u)
\]
is the probability for a neuron to be in the refractory period at time $t$.

So long as the fluctuating term in Eq.~(\ref{eq:FP}) can be neglected, the dynamics
relaxes towards a stationary state which can be interpreted as an asynchronous regime
characterized by a constant current $\nu_0$ and constant fluctuations $\sigma_0$, 
\begin{eqnarray}
\mu_0 &=& -c\tau (1-b)g_1 \mathcal{J}\nu_0 \\
\sigma^2_0 &=&   \frac{\tau \mathcal{J}^2}{1-b}\nu_0 \; ,
\label{eq:mu0sigma0}
\end{eqnarray}
which can be determined self-consistently using the following
expression for $\nu_0$,
\begin{equation}
\frac{1}{\nu_0} = \tau_r + \tau \sqrt{\pi} \int_{\frac{V_r-\mu_0-\mu_e}{\sigma_0}}^{\frac{V_{th}-\mu_0-\mu_e}{\sigma_0}}
du \ \mathrm{e}^{u^2} (1 + \mathrm{erf}(u))
\label{eq:nu}
\end{equation}

It is now convenient to introduce the following changes of variables
\[
Q = \frac{\sigma_0}{2\tau\nu_0}P \qquad y = \frac{v-\mu_0-\mu_e}{\nu_0} \qquad n = \frac{\nu}{\nu_0}-1 \; .
\]

The threshold and reset potentials become,
\[
y_\theta = \frac{v_\theta-\mu_0-\mu_e}{\sigma_0}  \qquad  y_r = \frac{v_r - \mu_0-\mu_e}{\sigma_0}
\]
while the Fokker-Planck equation can be rewritten as
\begin{eqnarray}
\tau \frac{\partial Q}{\partial t} &=& \frac{\partial }{\partial y}
\left (
y - n(t-\tau_d) \frac{\mu_0}{\sigma_0}
\right ) Q  + \nonumber \\
&& + \frac{1+n(t-\tau_d)}{2} \frac{\partial^2Q}{\partial y^2} 
+ \sqrt{c \tau } \frac{\partial Q}{\partial y}\zeta(t) \; ,
\label{eq:FPN}
\end{eqnarray}
accompanied by the boundary conditions
\[
\frac{\partial Q}{\partial y}(y_\theta) = -\frac{1+ n(t)}{1+n(t-\tau_d)}
\]
and
\[
\frac{\partial Q}{\partial y}(y_r^+) -\frac{\partial Q}{\partial y}(y_r^-)  =
\frac{1+ n(t-\tau_r)}{1+n(t-\tau_d)}
\]

The stationary solution can be expressed as
\begin{eqnarray*}
Q_0(y) &=& \mathrm{e}^{-y^2} F(y) \quad y>y_r  \\
Q_0(y) &=& \mathrm{e}^{-y^2} F(y_r) \quad y<y_r
\end{eqnarray*}
where
\[
F(y) = \int_y^{y_\theta} du \mathrm{e}^{u^2} \; .
\]
Finally, from the definition of $n$ it follows that $n_0 = 0$.

As a next step, we linearize the Fokker-Planck equation around the stationary
solution to treat the fluctuations in a perturbative way.
Upon assuming $Q = Q_0+q$ and neglecting nonlinear terms in $q$ and $n$,
\begin{eqnarray}
\tau \frac{\partial q}{\partial t} &=& \frac{\partial yq }{\partial y}
 + \frac{1}{2} \frac{\partial^2q}{\partial y^2} -
n(t-\tau_d) \left (\frac{\mu_0}{\sigma_0} \frac{dQ_0}{dy} -
\frac{1}{2} \frac{d^2Q_0}{dy^2} \right) \nonumber  \\
&& + \sqrt{c \tau } \frac{\partial q}{\partial y}\zeta(t) \; , 
\label{eq:FPL}
\end{eqnarray}
while the b.c. can be rewritten as

\[
\frac{\partial q}{\partial y}(y_\theta) = -1 -n(t) + n(t-\tau_d)
\]

and

\[
\frac{\partial q}{\partial y}(y_r^+) -\frac{\partial q}{\partial y}(y_r^-)  =
1+ n(t-\tau_r)-n(t-\tau_d)
\]
Eq.~(\ref{eq:FPL}) is a linear Langevin equation operating in an infinite-dimensional space.
The best way to handle it is to Fourier transform Eq.~(\ref{eq:FPL}), introducing ${\hat q}(y,\omega)$
and $\hat n$. This way we obtain two first order ode's for each frequency
variables,
\begin{eqnarray}
\frac{d\hat q}{dy} &=& {\hat u} \label{eq:FQ2} \\
\frac{d\hat u}{dy} &=& -2y {\hat u} + 2(i\omega\tau-1) {\hat q} + G(\hat n,y) \nonumber
\end{eqnarray}
where
\[
G(\hat n,y) = \mathrm{e}^{-i\omega \tau_d} {\hat n} \left ( \frac{2 \mu_0}{\nu_0} \frac{dQ_0}{dy}-
 \frac{d^2Q_0}{dy^2}\right) - 2\sqrt{c \tau}\frac{dQ_0}{dy}
\]
and we have implicitly assumed that the power spectrum of $\zeta(t)$ is flat and equal to 1.
The corresponding b.c. write as
\begin{equation}
\hat q (y_\theta,\omega)=0 \qquad {\hat u}(y_\theta) =\hat n(\mathrm{e}^{-i\omega \tau_d}-1)
\label{eq:bc1}
\end{equation}
and
\begin{equation}
{\hat u}(y_r^-) = {\hat u}(y_r^+) - {\hat n}[\mathrm{e}^{-i\omega \tau_d} -\mathrm{e}^{-i\omega \tau_r} ]
\label{eq:bc2}
\end{equation}
These equations have been numerically solved by integrating Eq.~(\ref{eq:FQ2}) 
backward in $y$, starting from $y=t_\theta$ for each given frequency $\omega$
and a tentative value of $\hat n(\omega)$, using Eq.~(\ref{eq:bc1}) to select the initial conditions
for $\hat q$ and $\hat u$.
The integration is then stopped at $y=y_r$, where the right derivative $\hat u(y_r^+)$ 
is adjusted according to Eq.~(\ref{eq:bc2}) to obtain the left derivative $\hat u(y_r^-)$
and thereby proceed towards $-\infty$. Only if the initial value of $\hat n$ is correct, 
$q(y)$ converges towards zero.

  

\end{document}